\documentclass[journal]{IEEEtran}
\usepackage{mathrsfs}
\usepackage[cmex10]{amsmath}
\usepackage{cite,graphicx,epsfig,wrapfig,amsfonts,amssymb,algorithmic,algorithm,threeparttable,url,subfigure}
\usepackage{mdwmath}
\usepackage{mdwtab}
\usepackage{amsthm}
\usepackage{hyperref}

\theoremstyle{definition}

\usepackage[T1]{fontenc}
\usepackage[USenglish,UKenglish,french,spanish,italian]{babel}
\usepackage[nodayofweek,level]{datetime}

\newcommand{\mydate}{\formatdate{10}{4}{2015}}

\graphicspath{{./}{figures/}}

\ifCLASSINFOpdf
\else
\fi
\begin{document}

\begin{titlepage}

\begin{tabular}{l        r}

\includegraphics[bb=20bp 00bp 500bp 450bp,clip,scale=0.3]{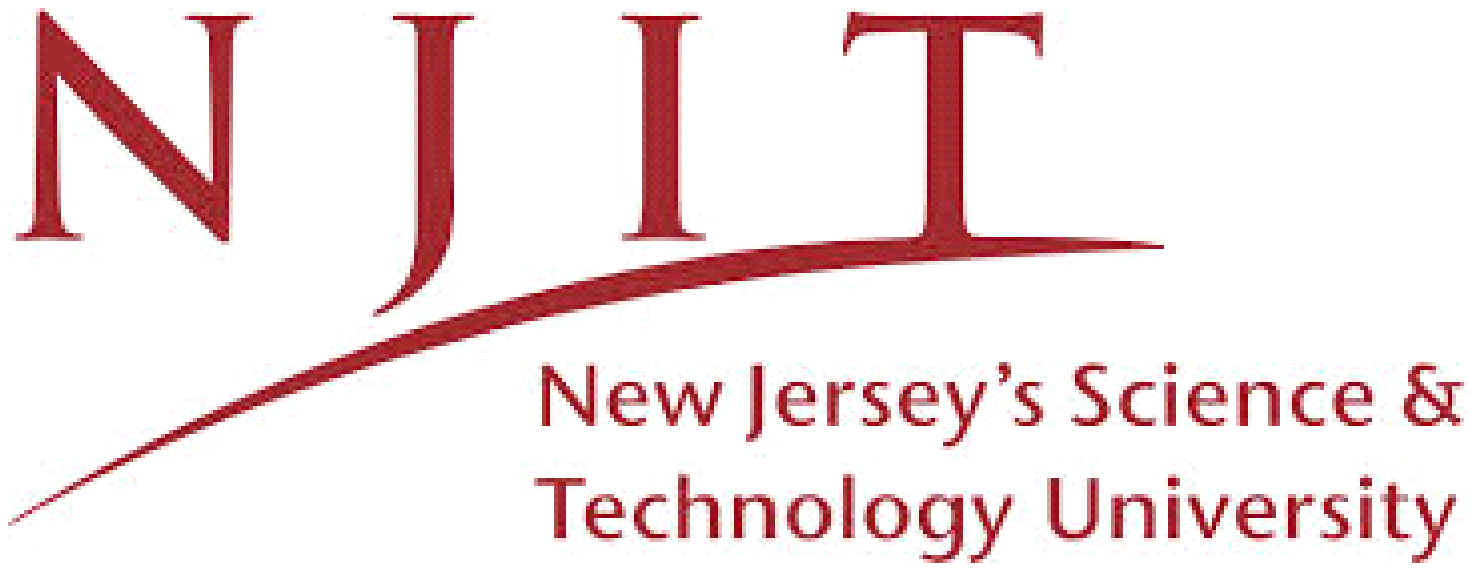} \hspace{6cm} & \includegraphics[bb=0bp -200bp 500bp 550bp,clip,scale=0.2]{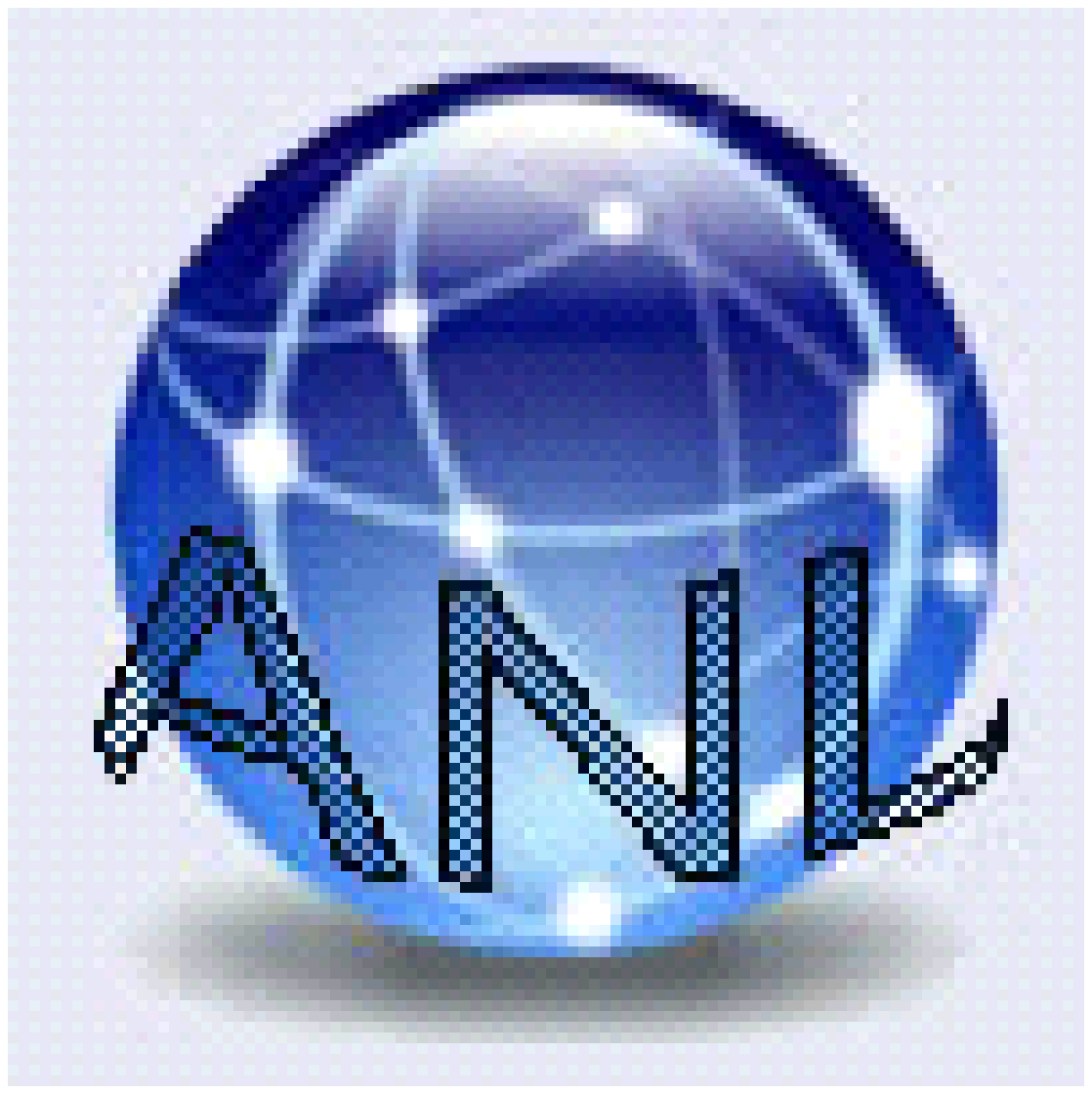}

\end{tabular}

\begin{center}

\textsc{\LARGE GATE: Greening At The Edge}\\[1.5cm]

{\Large \textsc{nirwan ansari}}\\ 
{\Large \textsc{tao han}}\\ 
{\Large \textsc{mina taheri}}\\[2cm]

{}
{\textsc{TR-ANL-2015-003}\\
\selectlanguage{USenglish}
\large \mydate} \\[3cm]

{\textsc{Advanced Networking Laboratory}}\\
{\textsc{Department of Electrical and Computer Engineering}}\\
{\textsc{New Jersy Institute of Technology}}\\[1.5cm]
\vfill

\end{center}

\end{titlepage}


\selectlanguage{USenglish}
\begin{abstract}
Dramatic data traffic growth, especially wireless data, is driving a significant surge in energy consumption in the last mile access of the telecommunications infrastructure. The growing energy consumption not only escalates the operators' operational expenditures (OPEX) but also leads to a significant rise of carbon footprints. Therefore, enhancing the energy efficiency of broadband access networks is becoming a necessity to bolster social, environmental, and economic sustainability. This article provides an overview on the design and optimization of energy efficient broadband access networks, analyzes the energy efficient design of passive optical networks, discusses the enabling technologies for next generation broadband wireless access networks, and elicits the emerging technologies for enhancing the energy efficiency of the last mile access of the network infrastructure.
\end{abstract}
\IEEEpeerreviewmaketitle

\section{Broadband Access Networks}
\begin{figure*}[ht!]
     \begin{center}
        \subfigure[Optical access network]{%
            \label{fig:opt}
            \includegraphics[width=2.5in]{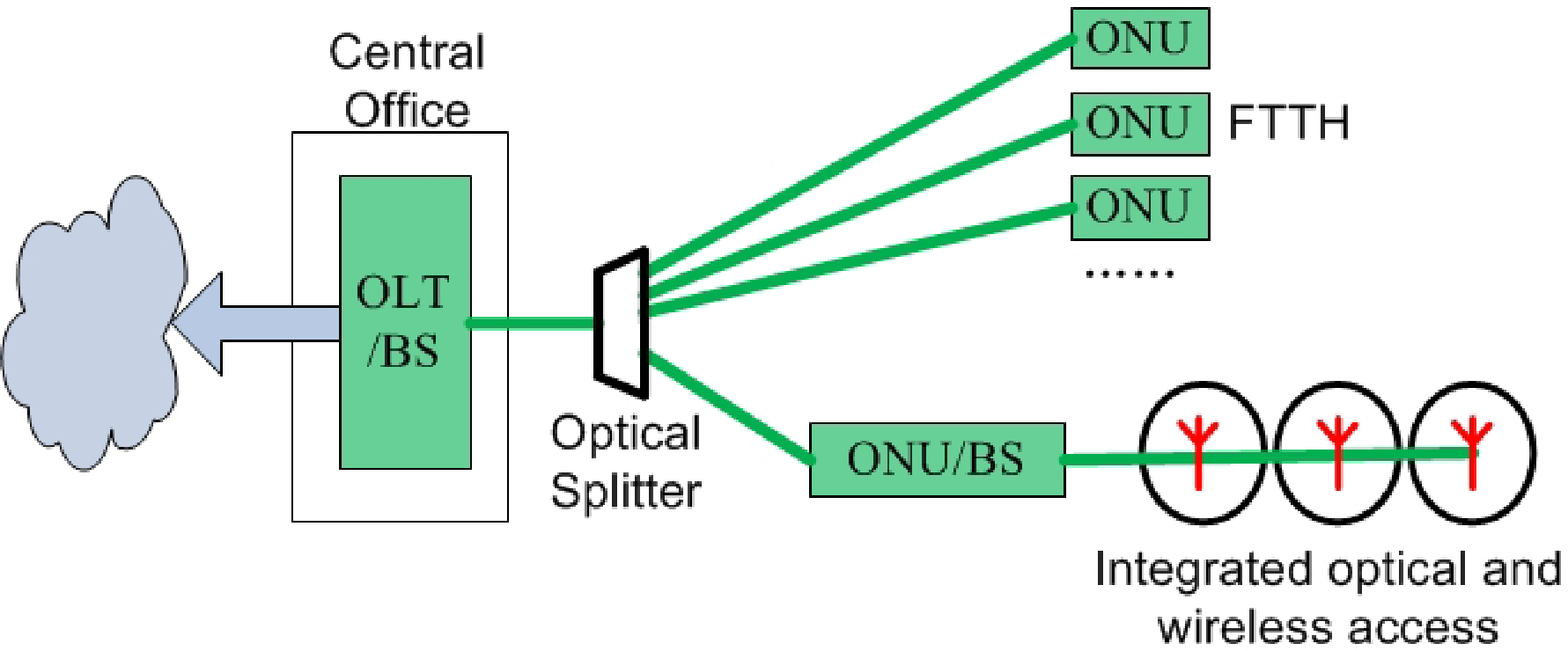}
        }%
        \subfigure[Wireless access network]{%
            \label{fig:wireless}
            \includegraphics[width=2.5in]{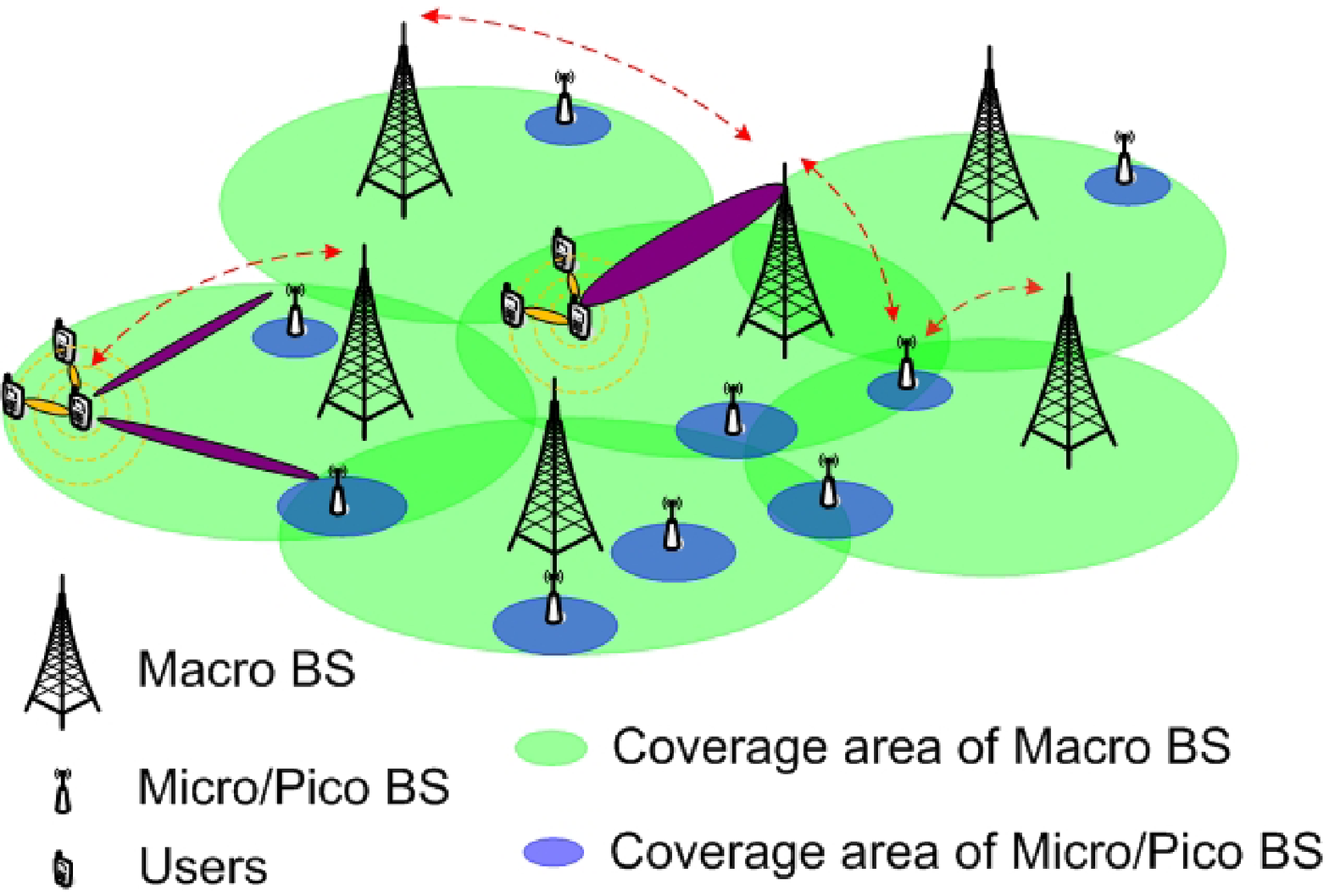}
        }%
     \end{center}
    \caption{%
      Prevailing choices of the last mile access.
     }%
   \label{fig:access}
\end{figure*}

Access networks are the last mile in the Internet access. As Internet traffic surges, broadband access networks should be upgraded to provide high network capacity. Currently, broadband access can be offered over the following communication media: digital subscriber line, hybrid fiber coaxial cable, broadband over powerline, wireless, and optical fiber. Owing to the immense capacity of optical fiber, optical access network will be extensively deployed for future broadband access. Furthermore, the ubiquity of mobile and wireless devices warrants wireless access networks an indispensable option for broadband access. Optical (Fig. \ref{fig:opt}) and wireless (Fig. \ref{fig:wireless}) access will be the predominant choice of the last mile access. GATE (Greening At The Edges) refers to the transformation of the access portion of communications infrastructure into an energy efficient version, i.e., an environmentally friendly version. Thus, the GATE's objective is to green both wireline and wireless access networks which currently consume a significant amount of energy of communications infrastructure owing to the large quantity of access nodes.

\begin{figure*}[ht!]
     \begin{center}
        \subfigure[The P2P Fiber access network]{%
            \label{fig:p2p_fiber}
            \includegraphics[width=1.8in]{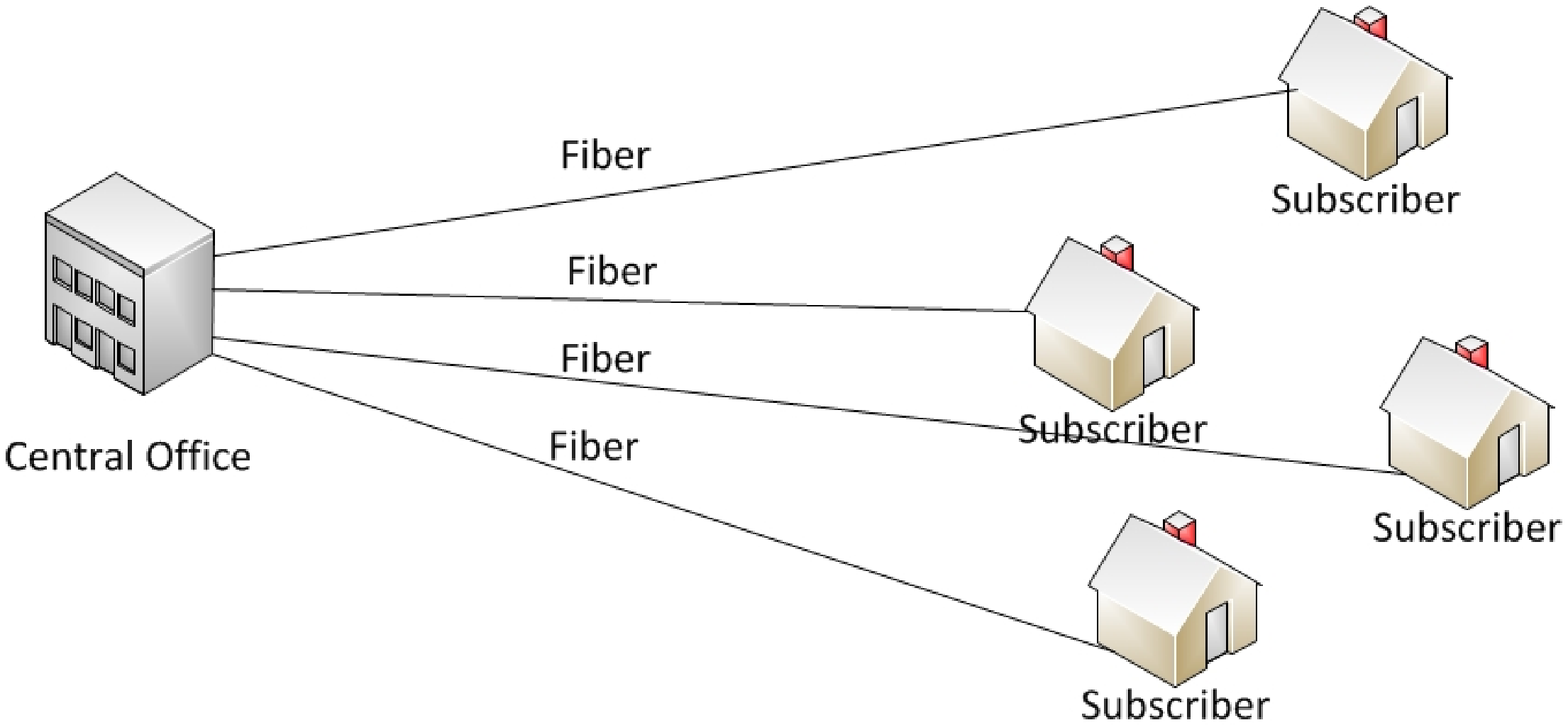}
        }%
        \subfigure[The Active Ethernet network]{%
            \label{fig:aen}
            \includegraphics[width=1.8in]{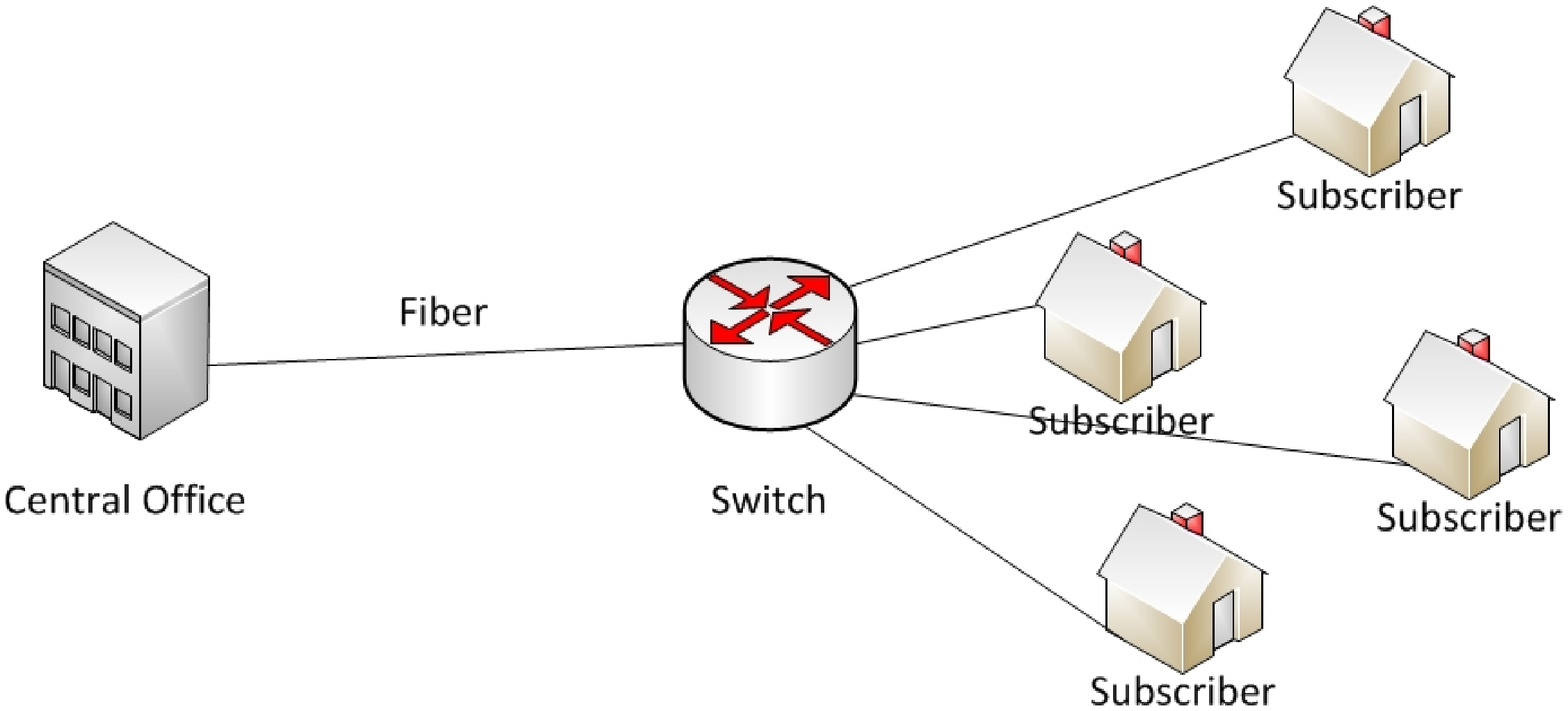}
        }%
        \subfigure[The Passive optical network.]{%
           \label{fig:pon}
           \includegraphics[width=1.8in]{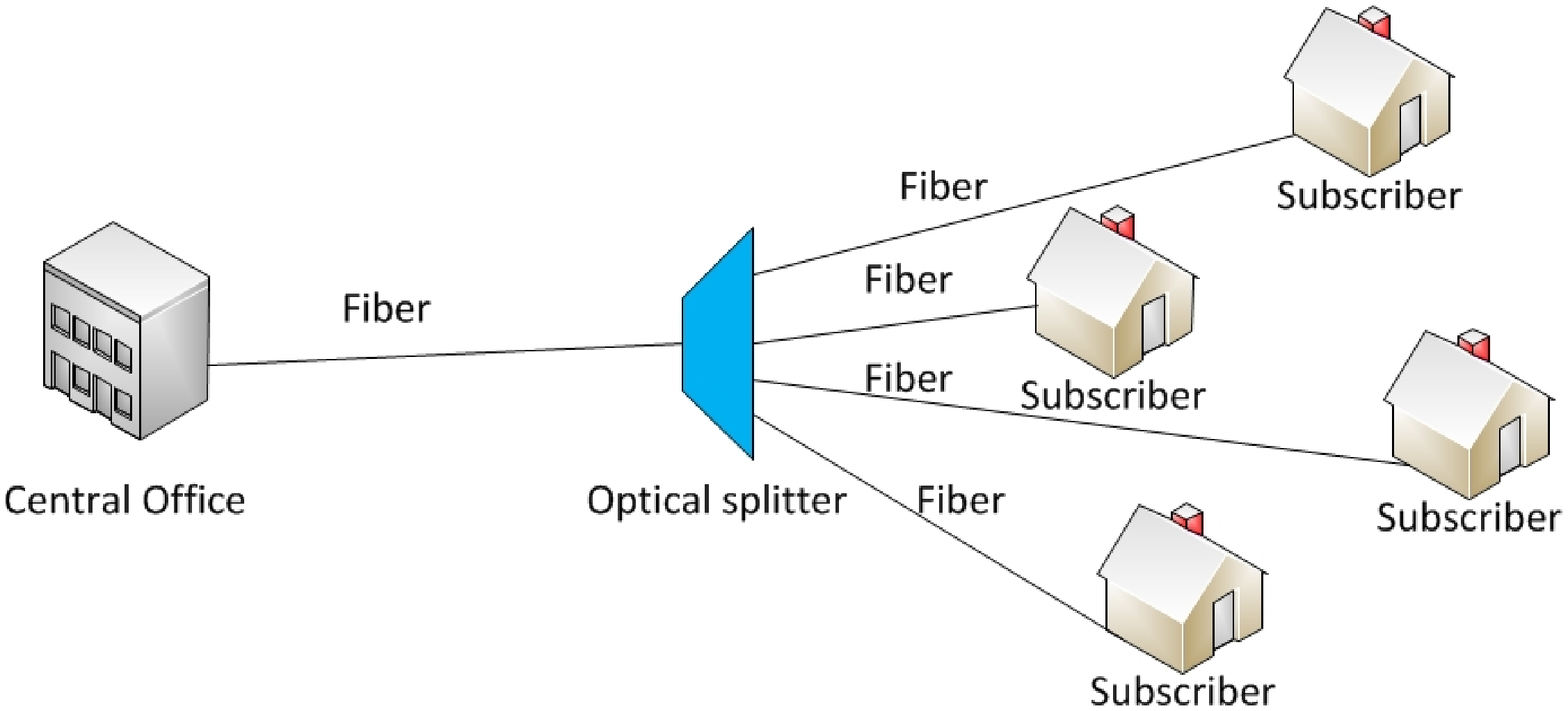 }
        }%
\end{center}
    \caption{%
      The optical access networks.
     }%
   \label{fig:optical_access}
\end{figure*}

\subsection{Optical Access Networks}
Optical access networks feed the metro and core networks by gathering data from end subscribers. In general, there are three major types of optical access networking technologies \cite{Ansari:2013:MAC}:
\begin{itemize}
\item \textit{Point-to-Point Fiber} connects fiber directly between the central office and the optical network terminals (ONTs) located at the subscribers' homes. Since the fiber is dedicated to individual subscribers, the optical power experiences small loss, and the transmission power budget allows a distance of up to 10 km between the central office and the subscribers' homes. Point-to-Point fiber access networks (Fig. \ref{fig:p2p_fiber}) has a simple network architecture and does not require expensive optical components.
\item \textit{Active Ethernet Network (AEN)} adopts a point to multiple points architecture as shown in Fig. \ref{fig:aen}. It utilizes electrical equipment, e.g., switches and 
routers, to distribute signals, thus enabling network operators to completely control their infrastructures and provision quality of service at different levels. The electrical equipment aggregates fiber delivered directly to subscribers and dramatically reduce the number of fibers terminated in the central office. AEN can dedicate each subscriber with a high data rate link, e.g., 1 Gb/s. 
\item \textit{Passive Optical Network (PON)} has a similar architecture as AEN. PON (Fig. \ref{fig:pon}) uses a passive optical splitter to aggregate the signal. In the downstream, the splitter divides the light signal sending from the central office and then broadcasts it to all optical network units (ONUs). In the upstream, the optical splitter aggregates the light signals coming from ONUs and transmit the aggregated signal to the optical line terminal (OLT) over fiber. The network is referred to as passive optical network because there is no optical repeaters or any active devices in the network. As compared to AEN, the optical splitter adopted in PON requires zero power consumption and maintenance. As a result, the cost of deploying PON is significantly reduced. PON is also an enabling option for radio-over-fiber for integrated optical and wireless access \cite{Zhang:2011:SAC}.

\end{itemize}
\subsection{Broadband Wireless Access Networks}
Broadband wireless access networks provide wireless data communications at a comparable data rates to that of wireline access networks for subscribers. In general, wireless access networks can be classified into three categories according to their coverage areas: wireless local area networks (WLANs), wireless metropolitan area networks (WMANs), and wireless wide area networks (WWANs). WLANs provide wireless data communications in an area with a cell radius of up to hundreds of meters. WMAN covers a wider area that can be as large as an entire city while WWAN provides wireless data services to an area covering multiple cities.

The most popular WLAN technology is WiFi which adopts IEEE 802.11 standards. The service radius of WiFi is usually from 50 to 100 meters. WiFi initially provides an aggregated throughput of 11 Mbps, which has been significantly enhanced by recent IEEE standards, e.g., IEEE 802.11ac. WiFi networks are usually self-deployed at home and office environment. Recently, carriers such as Comcast and AT\&T also deploy outdoor WiFi access points as hot spots to offloading traffic from mobile cellular networks \cite{Han:2014:EMT}. 

Mobile cellular networks are one of the major WWAN technologies. The network capacity of cellular networks has been significantly enhanced via the network evolution from GSM to LTE-Advanced, which can provide a peak data rate as high as 1 GB/s. The capacity of mobile cellular networks will continue to improve by adopting advanced technologies such as massive Multiple-Input-Multiple-Output (MIMO), coordinated multi-point (CoMP), small cell networks, and device to device communications. In addition, mobile cellular networks and WiFi networks are gradually integrated with each other to provision high capacity and low latency wireless access networks.

\section{Green Optical Access Networks}
\begin{figure*}[ht!]
     \begin{center}
       \subfigure[The upstream transmission]{%
            \label{fig:upstream}
            \includegraphics[width=2.5in]{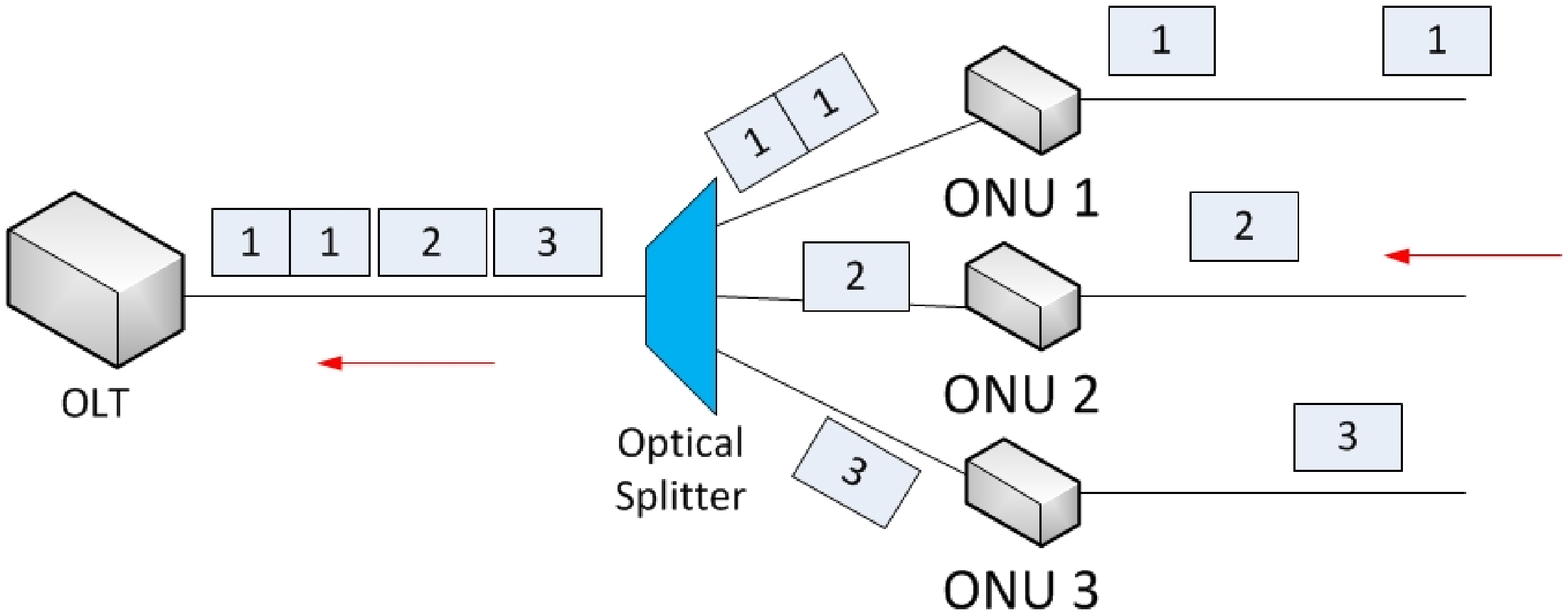}
        }%
        \subfigure[The downstream transmission.]{%
           \label{fig:downstream}
           \includegraphics[width=2.7in]{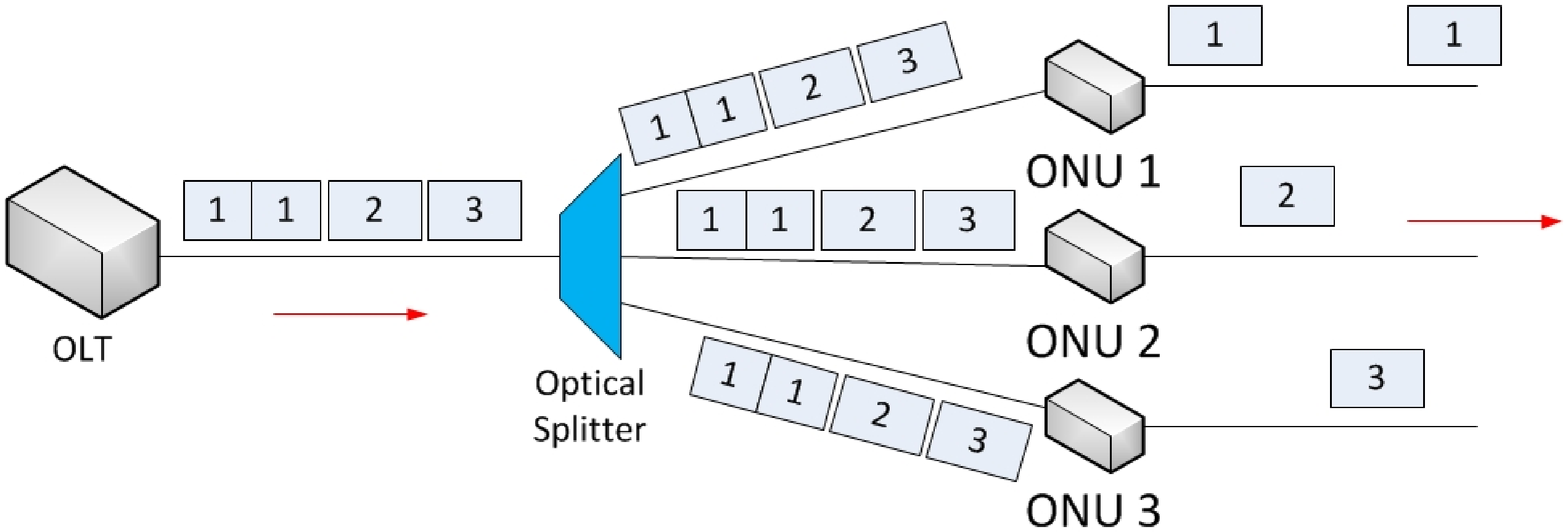 }
        }%
\end{center}
    \caption{%
      The TDM PON system.
     }%
   \label{fig:TMDPON}
\end{figure*}

Owing to the passive nature, PON consumes a small amount of energy per transmission bit. However, considering worldwide deployment of PON, it is worthwhile to further reduce the energy consumption of PON. In a PON system, e.g., TDM PON as shown in Fig. \ref{fig:TMDPON}, upstream traffic of ONUs is transmitted in the time slots assigned by OLT while the downstream traffic is broadcasted to all ONUs. The ONU and OLT are two key power consuming components in PON. Thus, enhancing the energy efficiency of PON is to reduce the energy consumption of ONUs and OLTs. 

\subsection{Reducing ONU energy consumption}
The energy efficiency of ONUs can be improved at both the physical and MAC layer. At the physical layer, research efforts are being made to develop low power optical transceivers and electronic circuits. At the MAC layer, scheduling and control schemes have been developed to enable ONU operating in multiple power modes for both upstream and downstream transmissions \cite{Zhang:2011:TEE}. With the support of multiple power modes, ONUs may disable certain functions and operate in low power mode to save energy when the traffic load is light. Since the downstream traffic is broadcased to all ONUs, it is challenging to dynamically switch ONUs to different power modes without interrupting network performance. To address this challenge, two major classes of schemes have been proposed. The first one is to design an MAC control scheme to convey the downstream queue status to ONUs to enable the ONU switching to the low power mode. For example, ONUs and OLT may perform two-way or three way handshakes to share the queue status and the power mode information. Typically, an OLT sends a control message to notify the downstream queue status to an ONU. Based on the queue status, the ONU may optionally enter different power modes, e.g., low power mode or sleep mode. The ONU will send its operating power mode back to the OLT. According to the ONU's operating power mode, the OLT will either buffer or transmit the arrival traffic toward the ONU.

The other class of schemes focus on investigating energy-efficient traffic scheduling and downstream bandwidth allocation. For example, a fixed bandwidth allocation (FBA) scheme can be implemented for downstream transmission when the network is lightly loaded. By implementing FBA, the time slots allocated to each ONU in each cycle are fixed and known by ONUs. Therefore, an ONU can switch to the sleep mode during time slots allocated to other ONUs. Recently, a simple and efficient sleep control scheme \cite{Zhang:2013:SCE}, which does not require handshaking between OLT and ONUs, has been demonstrated to achieve high bandwidth utilization and energy-saving efficiency. The basic idea to let each ONU infer its downstream queue status instead of being explicitly notified by the OLT.

\subsection{Reducing OLT energy consumption}

In PONs, OLT serves as the central access node to control the resource allocation of multiple ONUs. It is challenging to introduce multiple power modes into OLT for energy savings because the operation of OLT may affect the service of multiple ONUs. In the central office, one OLT chassis typically comprises multiple OLT line cards. To guarantee quality of service, the OLT line cards are usually powered on all the time. It is, however, possible to adapt the number of power-on OLT line cards according to the real-time arrival traffic to save energy. To avoid service degradation during the process of powering on/off OLT line cards, additional devices should be added into the legacy OLT chassis to facilitate communications of all ONUs with power-on line cards. 

\begin{figure}
\centering
\includegraphics[scale=0.6]{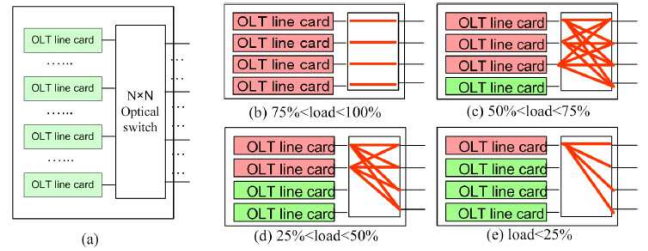}
\caption{The OLT with optical switch \cite{Ansari:2013:MAC}.}
\label{fig:olt_new}
\end{figure}

Fig. \ref{fig:olt_new}(a) shows a recently proposed novel OLT structure \cite{Ansari:2013:MAC} to dynamically switch on/off the OLT line cards, by placing an optical switch in front of all OLT line cards. The Optical switch dynamically configures the links between OLT line cards and ONUs. When the network is heavily loaded, the switch can be configured to dedicate one OLT line card to each PON system. When the network is lightly loaded, the switch can be configured such that multiple PON systems share one line card. In this case, some OLT line cards can be powered off to reduce the energy consumption. Since the process of changing the switch configuration is time consuming, frequent change of each line card's status may degrade the ONU performances. Therefore, the traffic of each PON is monitored for an observation period ($T_O$) defined as a multiple of the traffic cycle. If low channel utilization for a specific line card is detected, the line card can be powered off. On the other hand, to avoid service disruption, power-on OLT line cards should be able to provide the proper bandwidth for all the connected ONUs to the OLT chassis. Thus, traffic monitoring is performed for all the OLT line cards. If the total traffic remains less than a threshold during $T_O$, the line cards with the lower channel utilization can be powered off, and the ONUs communicating with these line cards will be connected to other power-on OLT line cards.  
 
Let $L$ be the total number of OLT line cards, $C$ the provisioned data rate of each line card, $N_j$ the total number of ONUs of the $j$th PON, and $R_{(i,j)}(t)$ the arrival traffic rate of the $i$th ONU of PON $j$ at time $t$. Defining the traffic load as $\mathop{\sum_{j=1}^L\sum_{i=1}^{N_j}}{R_{(i,j)}(t)}/(L \cdot C)$, the total number of line cards reduces to $x$ whenever the traffic load remains within the range [$(x-1)/L,x/L]$ in every traffic cycle of $T_O$. Since the traffic rate of each PON at time $t$ cannot exceed the fiber capacity, $\sum_{i=1}^{N_j}R_{(i,j)}(t) \leq C$ has to be satisfied. The power consumption of the optical switch is nonzero but far less than a single OLT line card. As a result, a significant amount of energy could be saved. 

Fig. \ref{fig:olt_new}(b)-(e) illustrates the switch configuration for the case that one OLT chassis contains four OLT line cards. By dynamically configuring the switch, the number of power-on OLT line cards is reduced from four to three, two, and one when the traffic load falls within the range $[50\%,75\%)$, $[25\%,50\%)$, and  $[0,25\%)$, respectively, during $T_O$.

By equipping the OLT chasis with proper optical switches, the communication between OLT line cards and ONUs can be dynamically configured. The new OLT structure appears to be promising and cost-effective as compared to the WDM based solutions.

\section{Green Wireless Access Networks}
Proliferation of wireless devices drive the exponential growth of wireless data traffic that leads to a continuous surge in capacity demands across wireless network infrastructures, notably the access portion. A wireless system is spectrum limited. In other words, the constraint of the available spectrum limits the capacity of a wireless system. Thus, various new technologies and network solutions have been proposed to enhance the capacity of wireless access networks. Since the premium goal is to provision high network capacity, these solutions usually focus on enhancing the spectrum efficiency. Thus, not all solutions are energy efficient. It is desirable to upgrade wireless access networks through energy efficient methods. In addition, providing larger network capacity usually incurs higher energy consumption. Therefore, network control and optimization methods are required to dynamically adjust the operation of wireless access networks to enable energy savings \cite{Han:2012:OGC}.
\subsection{Enabling technologies for high capacity green wireless access networks}
Massive Multiple-Input-Multiple-Output (MIMO)\cite{Larsson:2014:MMIMO}, Coordinated Multi-point (CoMP), Small Cell Networks (SCNs) \cite{Hoydis:2011:GSCN}, and Device to Device communications \cite{Doppler:2009:D2D} are emerging technologies to provision high capacity wireless access networks. These technologies not only improve the capacity but also potentially reduce the energy consumption of wireless access networks.  
\begin{enumerate}

\item \textit{Massive MIMO} utilizes an excessive number of antennas, e.g., hundreds of antennas, and can serve a large number of users simultaneously using the same time-frequency resources \cite{Larsson:2014:MMIMO}. Relying on simple phase-coherent processing of the signals, massive MIMO enables a significant network capacity enhancement as well as a dramatic radiate energy reduction. The network capacity enhancement comes from the aggressive spatial multiplexing exploited by massive MIMO technology. The energy efficiency improvement is achieved by focusing the radiate energy collected by the antennas into a small region where the targeted user is located.
\item \textit{CoMP} transmission is a promising technique to enhance the network capacity and energy efficiency for wireless access networks \cite{Barbieri:2012:CoMP}. By applying CoMP, multiple base stations (BSs) either jointly transmit data to mobile users or coordinately schedule their data transmissions. For example, multiple BSs may coordinate to align scheduling and beamforming to minimize the interference to users sharing the same time-frequency resource. In this case, coordinating scheduling and beamforming helps alleviate interference among users and BSs. Reducing interference may increase the capacity of wireless networks, and at the same time reduce the required radiate power.
\item \textit{Small Cell Networks (SCNs)} aim to reduce the distance between the radio transmitter, e.g., BSs, and the radio receiver, e.g., users. When the radio transmitter and receiver are close to each other, it requires less transmit power to compensate for the path loss and fading. As a result, the energy efficiency is improved. In addition, reducing the cell size enables more spectrum reuse and may significantly increase the spectrum efficiency \cite{Andrews:2014:AOLB}. Nowadays, small cell BSs are being deployed to offload traffic from macro BSs to enhance the energy and spectrum efficiency of wireless access networks. 
\item \textit{Device to Device Communications} is leveraged such that smart devices within proximity are able to connect with each other and form a communication network. Data traffic among the devices can be offloaded to the D2D network rather than delivering through BSs. For example, by enabling D2D communications, some users download contents from BSs while others may retrieve contents from their peers. In this way, D2D communications alleviates capacity demands in BSs and also reduces BS energy consumption \cite{Han:2014:OMT}. 
\end{enumerate}
\subsection{Multi-RAT traffic balancing}
Future wireless access networks will integrate multiple radio access technologies (RAT), e.g., LTE, SCNs, WiFi, and D2D communications. As a result, users will have a number of choices on selecting RAT to maximize their utilities. It is crucial to balance traffic loads among RAT to fully exploit the capacity of the heterogeneous wireless access network as well as to minimize the energy consumption of the network. In wireless access networks, balancing traffic load among BSs can be achieved via adapting user-BS associations. In general, optimizing user-BS associations in heterogeneous wireless access networks is a combinatorial optimization problem whose complexity grows exponentially with the scale of the network \cite{Andrews:2014:AOLB}.   

The simplest traffic balancing approach is the cell range expansion which biases the users' receiving power from some BSs to prioritize these BSs in associating with the users. For example, if BS A's bias is 5 dB while BS B's bias is 0 dB, a user will prioritize BS A in the process of the user associations. In this case, a user associates with BS B rather than BS A only when the user's perceived signal power from BS B is 5 dB more than that from BS A. The bias can be chosen based on various network optimization criteria. With respect to green communications, a large bias may be given to BSs with high energy efficiency. The cell range expansion approach is a practical solution. It is, however, difficult to find the optimal bias for the BSs to maximize the network utilities such as network capacity, traffic delivery latency, and energy consumption. 

In order to optimize traffic balancing, optimization theory has been applied to design the optimal user-BS association schemes. These approaches usually base on a simple traffic model and assume that users may partially associate with a BS. This relaxation enables low complexity algorithms to solve the user association problem. For example, Kim \emph{et al.} \cite{Kim:2012:DOU} assumed traffic arrived at a user follows Poisson distribution and designed a distributed user-association algorithm that maximizes the network utility. Han and Ansari \cite{Han:2014:PMN} also assumed the same traffic model and designed user-association with consideration of both network traffic delivery latency and renewable energy in BSs. The user-association algorithm, while optimizing the network utility, can achieve adaptable trade-off between the network energy consumption and the traffic delivery latency. 

The above traffic balancing schemes assume that users can freely associate with any wireless access points. However, this assumption does not always hold. For example, with strong revenue growth in wireless data markets, internet service providers (ISPs) such as Comcast and CableVision are densely deploying WiFi hot spots to provide WiFi connectivity to their customers in urban and suburban areas. However, these hot spots only open to their subscribers. In other words, without subscription, a user cannot associate with these hot spots and leverage them for data communications.  It is desirable to utilize hot spots deployed by ISPs to offload mobile data traffic. 

Since carrying mobile traffic introduces additional operation cost to ISPs, proper incentives should be given to ISPs for using their hot spots. Han and Ansari \cite{Han:2014:EMT} proposed a mobile traffic offloading scheme by leveraging cognitive radio techniques referred to
as energy spectrum trading (EST). EST exploits the merits of both mobile networks and ISP networks. Mobile networks are operating on licensed spectra that are not accessible by unlicensed users. Therefore, by proper spectrum management, mobile networks are able to provide their subscribers a variety of services with different QoS levels. However, as compared with the hot spots deployed by ISPs, BSs of mobile networks are usually sparsely deployed. Such deployments are not efficient in terms of the energy and spectral utilization.
On other hand, ISPs' hot spots are densely deployed, and are able to provide high speed data rates to their subscribers. However, operating on unlicensed spectrum, QoS of data services may not be guaranteed. EST enables mobile networks to offload data traffic to ISP networks to improve energy and spectral efficiency, and allows ISPs' hot spots access to the licensed spectrum to provide ISP data services with different QoS levels.
\begin{figure}
\centering
\includegraphics[bb=-150bp 100bp 680bp 780bp,clip,scale=0.3]{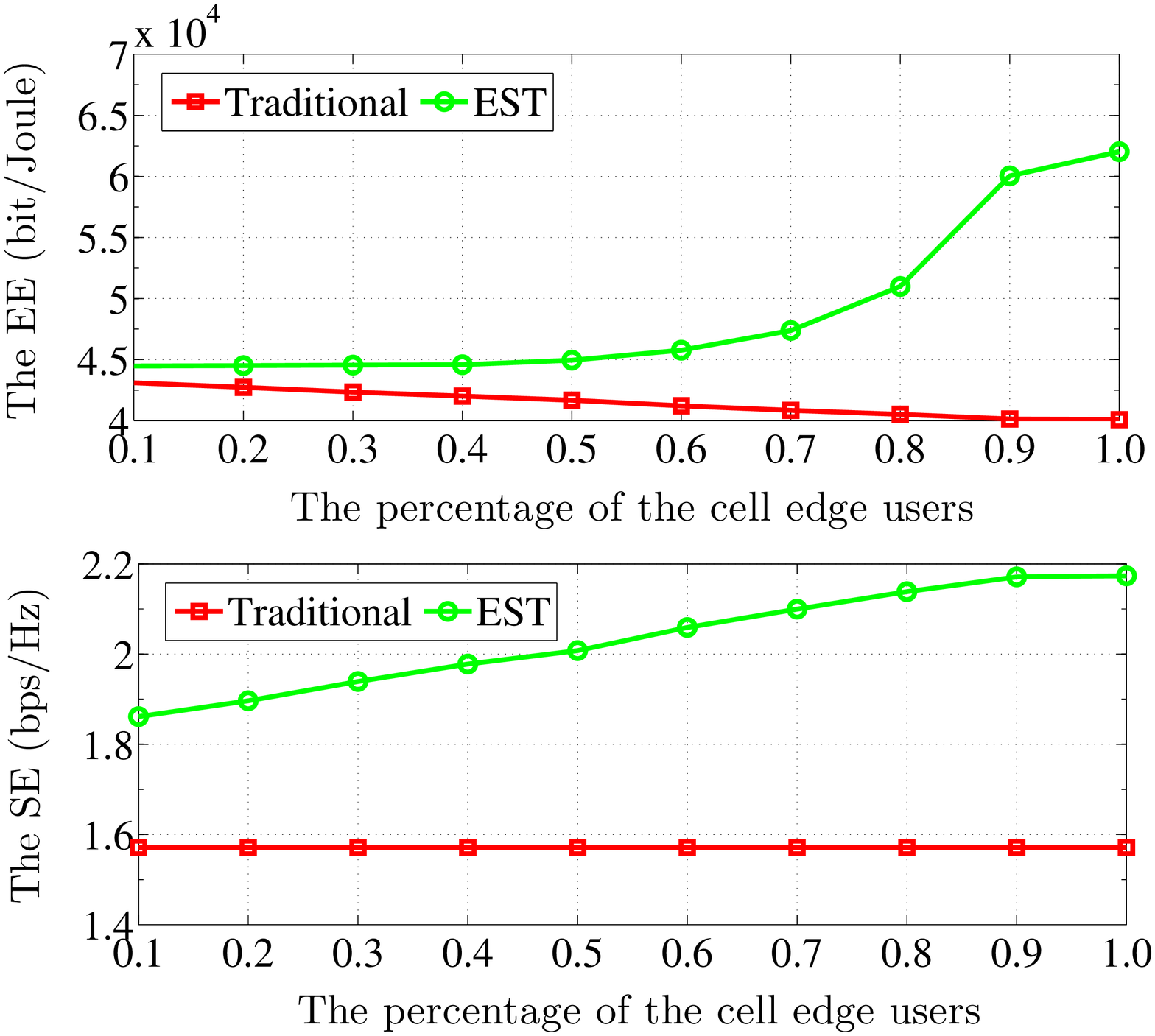} 
\caption{The performance of the EST scheme.}
\label{fig:fig_user_percentage}
\end{figure}
Fig. \ref{fig:fig_user_percentage} shows the energy efficiency (EE) and spectrum efficiency (SE) of the network versus the percentage of cell edge users. As the percentage of cell edge users increases, EE of the traditional scheme decreases because serving cell edge users usually requires more energy. EE of the EST scheme increases because more users are offloaded to IPSs' hot spots. For the same reason, SE of the EST scheme also increases.

\subsection{Wireless access network capacity adaptation}
Capacity demands of wireless access networks experience fluctuations. Meanwhile, a wireless access network operated at a high capacity usually consumes more energy than operated at a low capacity. For instance, when the capacity demand is decreasing, some BSs can be turned off to save energy. Thus, in order to enhance the energy efficiency of wireless access networks, it is desirable to adapt the capacity of wireless access networks based on traffic demands \cite{Niu:2011:TANGO}. 

The capacity demands for wireless access networks fluctuate because of two major reasons. The first one is the day-night behavior of users in terms of using wireless access networks. For example, users are usually more active in using wireless access networks during the day time than that during the night time. Therefore, the capacity demands for wireless access networks during the day are higher than those at night. The other reason leading to the variation in capacity demands is user mobility. Wireless connections allow users roam freely within the coverage area of the network. Since wireless access networks usually consist of multiple access points (APs)/BSs, the roaming of users leads to portion of the wireless access networks (some of the APs/BSs), where there are many active users, experience high capacity demand while the other portion of the networks has low capacity demands. Since wireless access points usually consume a significant amount of power even when it does not carry any traffic load, energy efficiency of a wireless access network is poor when the network is lightly loaded. To enhance the energy efficiency of wireless access networks, the network capacity should be adjusted based on traffic loads by dynamically switching on/off wireless access points.

In adapting the wireless access network capacity, user service requirements, e.g., minimum coverage and minimum quality of service, should be satisfied. In other words, when turning off some BSs, these BSs' coverage areas should be covered by their neighboring BSs. Otherwise, it will lead to high service blockage. Thus, in adapting the capacity of a wireless access network, coordination among BSs is necessary to ensure user QoS. Capacity adaptation is usually based on the estimated traffic load in individual BSs. When the traffic load in a BS is less than a threshold, the BS may be turned off to save energy. To avoid frequently switching BSs on/off, BSs should maintain their operating status (on or off) for a period of time. Therefore, traffic load estimation should at least represent the traffic load in the period of time. Hence, traffic load estimation should consist of three parts: traffic load from currently associated users, that from users who will hand over to other BSs, and that from users who will hand over to the current BS from neighboring BSs. The handover process requires the coordination of multiple BSs. Therefore, the coordination is crucial for the traffic load estimation in individual BSs to enable traffic load aware network capacity adaptation.

\subsection{Powering wireless access network with renewable energy}
\begin{figure}[t]
\centering
\includegraphics[scale=0.30]{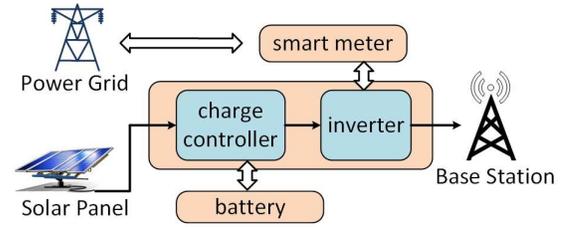}
\caption{A green energy powered BS.}
\label{fig:green_bs}
\end{figure}

Continuous advances in green energy technologies are improving the efficiency of generating electricity using renewable sources, e.g., solar, wind and sustainable biofuels. Meanwhile, the cost of deploying a green power system is driven down. It is promising to power wireless access networks with renewable energy to reduce grid energy consumption. Fig. \ref{fig:green_bs} shows a simplified diagram of a green energy powered BS \cite{Han:2014:PMN}. Five energy related components may be integrated into a BS to enable the utilization of renewable energy. These components are the green power generator, e.g., solar panel, the charge controller which regulates the output voltage of the green power generator, the DC-AC inverter, the battery, and the smart meter which enables power transmission between BSs and the power grid.

Define a green BS and a grid powered BS as the BS powered by renewable energy and grid energy, respectively. Although green BSs achieve zero carbon emission, the power supplies from renewable sources are not stable because the renewable energy generation highly depends on environmental
factors such as temperature, light intensity, and wind velocity. Thus, wireless access networks powered by renewal energy should be designed carefully to accommodate the dynamics of renewable sources.

A wireless access network is referred to as a green wireless access network when a portion of its APs/BSs can be powered by renewable energy. Based on power supplies, green wireless access networks can be classified into off-grid green wireless access networks whose BSs are powered by solely renewable energy, on-grid green wireless access networks whose BSs are powered by both renewable energy and on-grid energy, and hybrid green wireless access networks which consist of both green BSs and grid powered BSs. 

\begin{enumerate}
\item \textit{Off-grid green wireless access networks} aim to utilize the harvested energy to sustain traffic demands of users in the network \cite{Han:2014:PMN}. 
The optimal utilization of renewable energy involves two aspects: optimizing energy allocation in multiple time slots by determining how much energy should be used at the current time slot, and how much energy is reserved for future time slots for individual BSs; balancing traffic load among APs/BSs based on the availability of renewable energy. Owing to mobility, APs/BSs in different locations may experience various traffic loads. 
An AP/BS's power consumption is closely related to its traffic intensity, and thus the power consumption in APs/BSs may be different. In order to sustain traffic demands of all users, it is desirable to balance traffic load among BSs according to the status of green energy.

\item \textit{On-grid green wireless access networks} utilize renewable energy to reduce on-grid power consumption and grid power as a backup power source to compensate for the power demand which exceeds the amount of electricity generated from renewable energy. BSs connected to the main grid may not require energy storage as power backups. However, energy storage, if installed in individual BSs, can store renewable energy and help shift the peak grid power demand, thus reducing OPEX as well as alleviating the $CO_{2}$ emission. Since the electricity price is highly correlated to demands, the electricity price in peak hours is usually higher than that in off-peak power demand hours. In order to reduce OPEX, green energy is utilized when the electricity price is higher than a threshold, and is stored in the batteries when the electricity price is low. This also helps alleviate power congestion in power grid and thus reduces power loss in power grid.

In order to optimize the utilization of renewable energy, the design of on-grid green wireless networks should be aware of the status of renewable energy. Renewable energy aware user association \cite{Han:2014:PMN} is proposed to optimize both the network utilities and the renewable energy utilization. In addition, the renewable energy aware BS sleeping mode can be investigated to dynamically switch on/off BSs based on not only their traffic loads but also the availability of renewable energy.

\item \textit{Hybrid green wireless access networks} aim to minimize the on-grid power consumption and maximize the utilization of green energy by guiding more traffic loads to green BSs. One simple approach to offload data traffic to green BSs is to adjust the handover parameters to prioritize green BSs. This approach adjusts the handover parameters of BSs to enable wireless users to more easily handover to green BSs than to grid powered BSs.
Another approach is to increase the transmit power of the green BSs, thus enlarging the coverage areas of these BSs. As a result, more traffic will be offloaded to the green BSs \cite{Han:2014:PMN}.

\end{enumerate}
\section{Further Challenges to Overcome}	
According to Energy 2020\footnote{\url{http://www.scte.org/energy2020/}}, access network power supplies and edge facilities consume over 70\% of a cable operator's energy consumption. Therefore, substantial innovation is needed to further enhance energy efficiency in these areas, i.e., edges that house headend and hub equipment, especially OLT. While we have proposed a solution to reduce energy consumption of OLT (Fig. 4), fixed-rate OLT line cards limit further energy saving. To satisfy the QoS requirement, when traffic load is slightly higher than the capacity of a line card, the second line card needs to be activated. Digital signal processing can be utilized to build rate adaptive transceivers. Such software-defined transceivers can support a set of data rates without increasing the number of wavelength channels. Therefore, the OLT can update the wavelength data rate based on the traffic volume in a finer granularity and can thus potentially achieve further energy saving. Designing such capacity-adaptive optical transceivers remains a challenging issue in GATE.  
 
Furthermore, Fiber-Wireless (FiWi) networks are considered to be future-proof energy efficient broadband access networks by intelligently capitalizing the merits of both optical and wireless technologies. Combined energy saving strategies for hybrid optical-wireless networks could save a significant amount of energy in future broadband access networks. Generating electricity by using renewable sources to power edge facilities will certainly help green the edges. We have briefly discussed how to power wireless access networks by renewable source. A FiWi topology can provide several redundant paths for a packet to reach its destination. Designing energy-efficient algorithms to route packets through the FiWi networks, in which both wireless and optical nodes are powered by green energy, is another research issue to be tackled.

On optimizing the wireless access networks, the trade-off between the network performance, e.g., network capacity and traffic delivery latency, and the efficiency of energy including green energy should be carefully evaluated. Enhancing the energy performance usually compromises the network performance, which may degrade the user quality of experience. To further GATE, it is desirable to adapt the network and energy performance based on the user network activities, i.e., how the users use the network and which applications are running.

To satisfy the ever-increasing network capacity demands, an increasing number of small cell BSs will be deployed. In ultra-dense small cell networks,  it is very important to dynamically adapt the operating statuses (on/off) of BSs. By doing this, the network operators can efficiently reduce the energy consumption as well as alleviate the interference among BSs. However, because of the very large number of small cell BSs, even the solution that approximates the optimal solutions may be very difficult to derive. It is desirable to develop new algorithms to adaptively optimize the ultra-dense small cell networks.

\section{Conclusion}
This article discusses the design and optimization of energy efficient broadband access networks including both optical and wireless access networks in order to meet the objective of GATE. We have briefly discussed various broadband access technologies and focused on greening passive optical networks (PONs) and broadband wireless access networks. We have investigated the design of energy efficient ONUs and OLTs of PON. We have overviewed the enabling techniques for next generation wireless access networks and investigated the emerging technologies for reducing the energy consumption of the networks. Finally, we have elicited some remaining open research issues to be tackled in order to further GATE.

\bibliographystyle{IEEEtran}
\bibliography{mybib}

\end{document}